# Longitudinal gOSNR Monitoring by Receiver-side Digital Signal Processing in Multi-Span Optical Transmission System

Choloong Hahn, Junho Chang, Zhiping Jiang[*]

Ottawa Research Centre, Huawei Technologies Canada, 303 Terry Fox Drive, Ottawa, Canada, zhiping.jiang@huawei.com

**Abstract** *We propose the world's first longitudinal gOSNR estimation by using correlation template method at Rx, without any monitoring devices located in the middle of the link. The proposed method is experimentally demonstrated in a 12-span link with commercial transceiver.*

**Introduction**

The capability of a channel's longitudinal power profile estimation (PPE) using only the received signal at receiver [1] is a pleasant surprise to the fibre optical communication community. Since then, research in this area has flourished, with a wide variety of applications [2], from fibre loss anomaly [3], filter offset [4], fibre type identification [5], PDL localization [6], gain spectra and tilts monitoring [7], to MPI localization [8]. Two main types of PPE algorithms, namely correlation-based method (CM) and minimum-mean-square-error-based method (MMSE), have been widely studied on their theoretical limit [9], spatial resolution [10], accuracy optimization [11], and adoption in commercial products [12]. PPE is quickly becoming a versatile tool for network operation and maintenance.

On the other hand, reflecting on the history, one may also be surprised by this question: why was PPE only proposed in 2019, more than 10 years after the idea of digital pre-compensation [13] was proposed and the receiver-based digital back propagation had been extensively investigated for fibre nonlinearity compensation. The answer may be the lack of recognition that the nonlinear noise could ever serve a beneficial role. Similarly, despite the numerous studies and applications on PPE, the principles still remain focused on utilizing the nonlinear distortions from the transmitted signal only. The link noise has been regarded as a mere nuisance, degrading PPE's monitoring quality. However, considering the link noise is the most important factor limiting the transmission quality, and there is currently no cost-effective way to monitor it, one may dare to ask an intriguing question: is it possible to monitor the link noise in a distributed manner by the receiver-side DSP only? While this task may seem highly challenging, if it can be achieved, the impact on the community would be profound. It enables a broader scope of transmission quality monitoring including the generalized optical signal-to-noise ratio (gOSNR), locations of noise sources, and noise figure of amplifiers over entire optical network, all without the necessity of deploying expensive devices, such as optical performance monitor (OPM) modules.

In this work, we propose utilizing the noise inherent in the received waveform to extract additional information beyond what conventional PPEs offer. Similar to the signal waveform, the noise waveform also introduces unique nonlinear distortions as it propagates along the link. By generating correlation templates from the received noise for CM, we demonstrate longitudinal gOSNR monitoring for the first time to our knowledge. It is experimentally confirmed in a 12-span optical link with commercial transceivers (TRx). By this work, we demonstrate that the noise is not just mere degrading factor in signal quality, but rather, it could turn into a valuable resource for signal quality monitoring.

**Principle of correlation-template based PPE**

Generally, PPE relies on the uniqueness of local waveform generated as a result of dispersion, and consequently unique Kerr nonlinear distortion which is recorded on the signal waveform as a part of noise. Fig. 1 shows the block diagram of a PPE method using correlation templates [8,12]. We may assume the linear waveform at Tx ($z = 0$) is $E_{\text{Lin}}(0,t) = \sqrt{P(0)}u(0,t) + \sqrt{N(0)}n(0,t)$, where $P(z)$, $N(z)$, $u(z,t)$, and $n(z,t)$ are the signal power, noise power at location $z$, normalized signal waveform, and normalized noise waveform at location $z$ and time $t$, respectively. In the enhanced regular perturbation (eRP) model, the nonlinear distortion at Rx ($z = L$) is $E_{\text{NL}}(L,t) \equiv \int_0^L \gamma P(z)\Delta u_z(L,t)dz$, where $\gamma$ is the nonlinear coefficient. The partial nonlinear distortion can be written as [9],

$$\Delta u_z(L,t) \equiv -j\widehat{D}_{z,L}\widehat{N}_{\text{eRP}}\widehat{D}_{0,z}E_{\text{Lin}}(0,t), \quad (1)$$

where $\widehat{D}_{z_1,z_2}$ is the chromatic dispersion (CD) operator corresponding to the distance from $z_1$ to $z_2$, $\widehat{N}_{\text{eRP}} \equiv (|\cdot|^2 - 2\langle|\cdot|^2\rangle)(\cdot)$ is the nonlinear operator based on eRP model, and $\langle\cdot\rangle$ is the time average. Thus, total waveform at Rx is

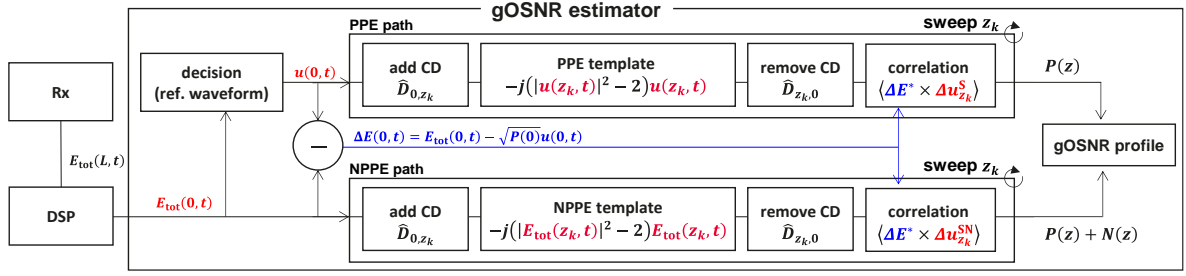

**Fig. 1:** Block diagram of correlation-template-based longitudinal gOSNR estimation

$E_{\text{tot}}(L,t) = E_{\text{Lin}}(L,t) + E_{\text{NL}}(L,t)$.

As shown in Fig. 1, the gOSNR estimator includes conventional PPE in the top path which uses a correlation template, $\Delta u_{z_k}^S(0,t) = -j\hat{D}_{z,L}\hat{N}_{\text{eRP}}\hat{D}_{0,z}u(0,t)$, generated from normalized signal waveform, $u(0,t)$ and Eq. (1). Since the template represents only the nonlinear distortion of the pure signal, it exclusively detects the local signal power.

**Local noise inclusive power monitoring**

In a similar manner to conventional PPE (e.g., signal PPE), having a template for the local noise waveform allows us to estimate the local noise power as well. However, the pure noise waveform, let alone the local noise waveform, is inaccessible with current technology. Instead of either pure signal or noise waveform, we use the total received waveform, $E_{\text{tot}}(0,t)$, to generate a correlation template, $\Delta u_{z_k}^{SN}(0,t)$, as illustrated in the bottom path of Fig. 1. The template for noise inclusive power profile estimation (NPPE) includes the nonlinear distortions from both the signal and noise waveforms, thus the NPPE can reveal the total power, $P(z) + N(z)$. Note that the noise power, $N(z)$, includes all existing noises such as amplified spontaneous emission (ASE), self-phase modulation (SPM), cross-phase modulation (XPM), and TRx implementation noise.

We perform a simulation for single channel 68-Gbaud dual-polarization (DP) QPSK signals propagating along a 6x80 km SSMF link, employing the split-step Fourier method. Here, the SPM noise is minimized by setting low signal launch power ($P_{\text{in}}$ = 0 dBm) for clear observation of the ASE noise detection. Firstly, a single-point ASE noise (set-SNR = 20 dB) is injected at 2nd, 4th, and 6th span input. NPPE performed at Rx for each span input location, and the profiles are plotted in Fig. 2(a). Here, $\text{SNR}(z) = P(z)/N(z)$ is the signal-to-noise ratio, the y-axis of the plot is NPPE correlation in dB normalized by signal power $(= (P(z) + N(z))/P(z))$. In Fig. 2(a), the NPPE increases after the injected location (vertical arrow) indicates that the noise power is detected on top of the signal power. Next, ASE noises with varying power levels are injected at 4th span and their NPPE profiles are shown in Fig. 2(b). NPPE increases are observed at injection location (vertical dashed line) corresponding to the ASE noise power. Finally, the ASE noise is injected at the input of every span (set-SNR = 20 dB for each injection) to mimic an amplifier chain (Fig. 2(c)), NPPE shows the increase of accumulated ASE noise power whereas PPE indicates the constant signal power.

**Longitudinal gOSNR estimation**

Since NPPE detects the sum of signal and noise power, the gOSNR can be calculated by using both PPE and NPPE as:

$$\text{gOSNR}(z_k) \equiv R \times \left(\frac{\langle \Delta E^*(0,t) \times \Delta u_{z_k}^{SN}(0,t)\rangle}{\langle \Delta E^*(0,t) \times \Delta u_{z_k}^{S}(0,t)\rangle} - 1\right)^{-1} \quad (2)$$

where $\Delta E(0,t) = E_{\text{tot}}(0,t) - \sqrt{P(0)}u(0,t)$ is the total noise waveform, $R = f_{\text{baud}}/12.5 \text{ GHz}$ is the bandwidth conversion factor from SNR to OSNR, and $f_{\text{baud}}$ is the signal baud rate. Here the ratio of NPPE and PPE is related to gOSNR because the noise power detected by NPPE is the total noise power in the optical link. A longitudinal gOSNR can be obtained by repeating the gOSNR estimation in Eq. (2) for all the locations along the link by sweeping $z_k$.

It is worth to note that the PPE and NPPE based on correlation template methods do not directly provide absolute power levels. The outcomes are the true powers convoluted with spatial response functions [9], and we may need proper scaling factors to obtain the true powers. However, the scaling factors are the same for both PPE and NPPE, and they can be cancelled out when we take the ratio in Eq. (2). This implies the proposed gOSNR estimation does not require any further scaling or correction.

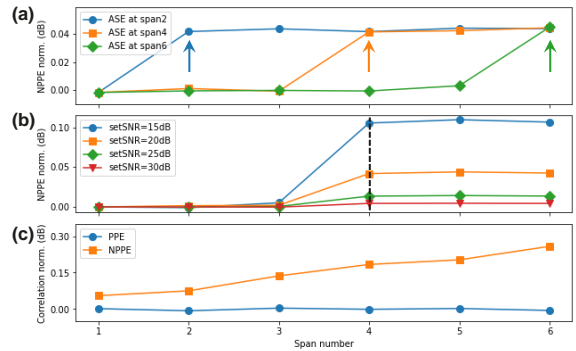

**Fig. 2:** Estimated NPPE profiles (a) Single-point ASE noise injection at different span (b) Single-point ASE noise injection with different power (c) ASE noise injection at every span.

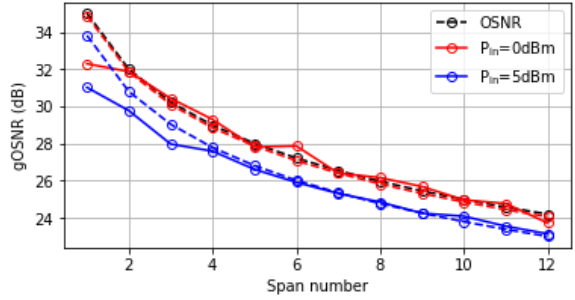

**Fig. 3:** Simulated longitudinal gOSNR over 12-span link with different signal launch power levels (solid) with their theoretical profiles (dashed).

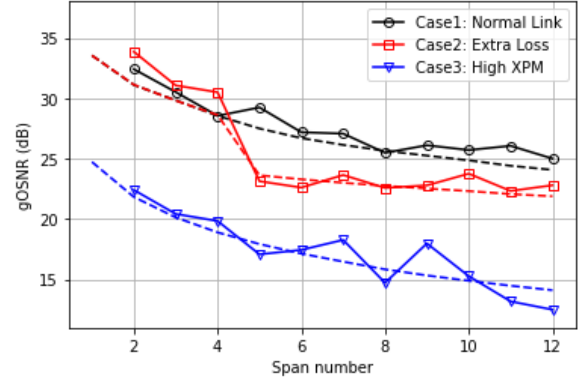

**Fig. 4:** Experimental estimation of longitudinal gOSNR over 12-span link with different cases. Case1: typical link with an EDFA at the input of each span (black). Case2: additional point-loss at the end of 4th span (red). Case3: enhanced XPM by increasing neighboring WDM channel power (blue).

The gOSNR monitoring method is firstly verified by simulation for 12x80 km SSMF link, with an amplifier chain. The simulation is conducted with a single channel 68-Gbaud DP-QKSP signal. Fig. 3 shows estimated gOSNR profiles for different $P_{in}$ with their theoretical gOSNR profiles (dashed). The black dashed curve is the set OSNR (i.e., only ASE noise) in the simulation, while the red and blue dashed curves indicate gOSNR, which also includes SPM noise corresponding to $P_{in}$. At $P_{in}$ = 0 dBm (red), the gOSNR is close to the set OSNR curve because the SPM noise is significantly lower than ASE noise. By increasing the signal power to 5 dBm, the gOSNR drop is observed due to the increased SPM noise power.

**Experimental demonstration of gOSNR estimation using commercial transceivers**

To demonstrate the longitudinal gOSNR estimation technique using a commercial product, we used a transceiver linecard supporting a 68-Gbaud DP-QPSK. The signal, together with 20 neighbouring WDM channels, was transmitted over the dispersion-uncompensated 12-span link consisting of 75-km long SMFs as a test channel. After transmission, the received signal was captured into the onboard memory and then processed with offline DSPs for data recovery. The demodulated symbols were then fed into gOSNR estimator shown in Fig. 1. The estimation process was repeated multiple times to reduce the inaccuracy by averaging. Fig. 4 shows the estimated gOSNR profiles (solid curves). Their theoretical expectations, obtained by the measured OSNR, and simulated SPM and XPM noises, are also shown in dashed curves. There were three test cases: 1) a link with an EDFA at the input of each span and negligible XPM (black curves), 2) the link with an additional point-loss of 7 dB at the end of 4th span which caused the gOSNR drop at 5th span and after (red curves), and 3) the link with enhanced XPM. In all cases, the launch power of the test channel at each span was maintained at a constant 5 dBm, while the neighbouring channels had power levels of -10 dBm for Case 1 and 2, and 8 dBm for Case 3.

The monitored includes all noises, but typically we can assume that the major contributions are ASE and nonlinear noises, the TRx implementation noise is much smaller, and can be calibrated out. Thus, the monitored is dominantly gOSNR. Case 2 represents the gOSNR changes by an exceptional loss in the middle of link (i.e. unexpected ASE noise), and Case 3 shows the gOSNR drop by XPM noise increase relative to Case 1. In all test cases, the estimated gOSNR profiles were in a good agreement with the calculated values.

**Conclusions**

We have presented the first work that leverages the noise within the received signal to monitor local noise power along the entire link. Longitudinal gOSNR monitoring by using two correlations, NPPE and PPE, is demonstrated in a 12-span link. Although our demonstration is based on correlation method, the idea may be easily adopted to other methods. Moreover, the demonstration was conducted with commercial transceiver that infers this method can be readily adopted in the field. In the future disaggregated systems, the distributed gOSNR monitoring capability by Rx only would greatly enhance the quality of transmission monitoring and troubleshooting. We believe that this work may stimulate new researches on the monitoring of optical link, and deliver various functionalities to the optical communication industry.